\documentstyle[emulateapj]{article}

\def\wig#1{\mathrel{\hbox{\hbox to 0pt{%
          \lower.5ex\hbox{$\sim$}\hss}\raise.4ex\hbox{$#1$}}}}

\def\etal{{\it et~al.\,}}
\def\mo{$M_\odot$}

\def\mj{$M_{\rm J}\,$}

\def\teff{T$_{\rm eff}\,$}
\def\teffs{T$_{\rm eff}$s$\,$}
\def\Dwa{$\,$\uppercase\expandafter{\romannumeral5}$\,$}
\def\mic{$\mu$m$\,$}

\def\sles{\lower2pt\hbox{$\buildrel {\scriptstyle <}
   \over {\scriptstyle\sim}$}}
\def\sgreat{\lower2pt\hbox{$\buildrel {\scriptstyle >}
   \over {\scriptstyle\sim}$}}
\def\sharpnull#1{}
\def\aa{Astron. Astrophys.\ }

\begin{document}

\slugcomment{\bf}
\slugcomment {Accepted to Ap.J.}

\title{The Near-Infrared and Optical Spectra of Methane Dwarfs and Brown Dwarfs}

\author{Adam Burrows\altaffilmark{1}, M.S. Marley\altaffilmark{2}, \& C.M. Sharp\altaffilmark{1}}

\altaffiltext{1}{Department of Astronomy and Steward Observatory, 
                 The University of Arizona, Tucson, AZ \ 85721}
\altaffiltext{2}{Department of Astronomy, New Mexico State University,
                 Box 30001/Dept. 4500, Las Cruces NM 88003}

\begin{abstract}

We identify the pressure--broadened red wings of the saturated potassium resonance lines at 7700 \AA\ as the source
of anomalous absorption seen in the near-infrared spectra of Gliese 229B and, by extension,
of methane dwarfs in general.  In broad outline, this conclusion is supported by the recent work
of Tsuji {\it et al.} 1999.  The WFPC2 $I$ band measurement of Gliese 229B is also consistent with this hypothesis.
Furthermore, a combination of the blue wings of this K I resonance doublet, the red wings of the Na D lines at 5890 \AA,
and, perhaps, the Li I line at 6708 \AA\ can explain in a natural way the observed WFPC2 $R$ band flux of Gliese 229B.
Hence, we conclude that the neutral alkali metals play a central role in the near-infrared and optical
spectra of methane dwarfs and that their lines have the potential to provide crucial diagnostics
of brown dwarf properties.  

The slope of the spectrum from 0.8 \mic to 0.9 \mic for the Sloan methane dwarf, SDSS 1624+00, 
is shallower than that for Gliese 229B and its Cs lines 
are weaker. From this, we conclude that its atmosphere is tied to a lower core entropy
or that its K and Cs abundances are smaller, with a preference for the former hypothesis.  
We speculate on the systematics of the near-infrared and optical spectra of methane dwarfs,
for a given mass and composition, that stems from the progressive burial with decreasing \teff 
of the alkali metal atoms to larger pressures and depths.   

Moreover, we surmise that those extrasolar giant planets (EGPs) that achieve
\teffs in the 800--1300 K range due to stellar insolation will show signatures of the
neutral alkali metals in their albedo and reflection spectra.  We estimate that, 
due predominantly to absorption by Na D lines, the geometric
albedo of the EGP $\tau$ Boo b at $\lambda=0.48$ \mic is $<0.1$, consistent 
with the new (and low) upper limit of $0.3$ recently obtained by Charbonneau \etal (1999).

\end{abstract}

\keywords{brown dwarfs, Gliese 229B, SDSS 1624+00, methane dwarfs, T dwarfs, 
extrasolar giant planets, alkali metals, atmospheres, non--gray spectral synthesis}

\section{Introduction}

The discovery of Gliese 229B in 1995 was a milestone in the study of brown dwarfs,
providing the first bona fide object with an effective temperature (T$_{\rm eff} \sim$ 950 K) that 
was unambiguously substellar (Nakajima \etal 1995; Oppenheimer \etal 1995).  
It also validated in dramatic fashion the allied theoretical and observational efforts
of the many groups around the world that were engaged in brown dwarf 
research and started a race to discover and understand brown dwarfs  
that shows no sign of abating.  Indeed, {\it seven} similar substellar mass objects, the so-called ``T" or ``methane" dwarfs,
have since been discovered in the field by the Sloan Digital Sky Survey (Strauss \etal 1999; Tsvetanov \etal 1999),
by the 2MASS survey (Burgasser \etal 1999), and by the NTT/VLT (Cuby \etal 1999).  Superficially,
these all seem to be clones of Gl 229B, but there are subtle spectral differences that no doubt reflect true differences
in mass, metallicity, and effective temperature.  In addition to this flurry of 
brown dwarf discoveries, the first new stellar spectral class in almost 100 years, the ``L type," has been defined,
with $\sim$100 members having been found to date by 2MASS (Kirkpatrick \etal 1999) and DENIS (Delfosse \etal 1997; Tinney \etal 1998).
These objects are characterized by the clear onset of refractory heavy metal depletion, in particular that of titanium and vanadium,
and by the growth in strength of the alkali metal lines of cesium, potassium, rubidium, and sodium
in their optical and near-infrared spectra.  L dwarfs also show distinct bands of FeH and CrH.   
Such transitions in chemical makeup with decreasing \teff are in keeping with theoretical  
predictions of the molecular consistuents of substellar atmospheres (Burrows and Sharp 1999; Fegley and Lodders 1996) that 
can serve to anchor and define the sequence of L dwarf spectral subtypes (Kirkpatrick \etal 1999).
The L dwarfs have \teffs from $\sim$1500 K to $\sim$2000 K and constitute the spectroscopic
link between M dwarfs and Gl 229B-like objects (Kirkpatrick \etal 1999).
Many L dwarfs are also brown dwarfs (with masses below the stellar boundary at $\sim$0.075 \mo),
but some will turn out to be stars very near the stellar edge.  The current ambiguity reflects 
the newness of this subject.

In this paper, we address T (methane) dwarf spectra in the optical and near infrared, focussing predominantly on Gl 229B
as the prototype.  Marley \etal (1996) and Allard \etal (1996) analyzed the full 
spectrum of Gl 229B (Oppenheimer \etal 1995; Geballe \etal 1996; Oppenheimer \etal 1998; 
Schultz \etal 1998) and obtained reasonable fits from the $J$ band 
($\sim1.2$ \mic) through the $N$ band ($\sim10$ \mic).  They concluded that its 
\teff was $\sim$900--1000 K and that its gravity ($g$) was $\sim 3\times10^4 - 10^5$ cm s$^{-2}$.  The gravity 
error bars were large and translated into a factor of $\sim$2--3 uncertainty in its inferred mass and age.  
They also concluded that the atmosphere of Gl 229B is indeed depleted in the refractory heavy elements,
such as Al, Si, Ca, Fe, and Mg, just as the theory of condensate formation 
and rainout in cool molecular atmospheres would suggest, and that its spectrum longward of $\sim$1 \mic
can be fit, in the main, by a simple mixture of H$_2$O, CH$_4$, and H$_2$.  
(Noll, Geballe, and Marley (1997) have detected CO at 4.67 \mic in Gl 229B as well.)
However, neither Marley \etal (1996) nor Allard \etal (1996) were able to fit the near-infrared 
observations between 0.8 \mic and 1.0 \mic and the theoretical excesses in flux 
ranged from 10 to 100.
%
%
The line marked ``Clear" in Figure 1 demonstrates the typical discrepancy between theory, that otherwise fits
rather well at longer wavelengths, and the observed spectrum of Gl 229B (Leggett \etal 1999).
Clearly, some ingredient is absorbing in the blue,
creating a very steep spectrum with a deep red (infrared) cast.  

Attempts to fit this problematic spectral interval were made by
Golimowski \etal (1998), who invoked a Tsuji \etal (1996) spectrum that retained TiO in the atmosphere   
to take full advantage of TiO's strong absorption in the blue, and Griffith \etal (1998), who hypothesized
that a population of small photochemical haze particles analogous to the red Titan Tholins (Khare and Sagan 1984) 
resided between $\sim$1400 K and $\sim$1800 K in Gl 229B's atmosphere.  Indeed, TiO does give roughly the
correct continuum slope, but it imposes the characteristic TiO bands on the spectrum that are not seen
in either late L or T dwarfs.  Furthermore, chemical abundance studies show that Ti and V are depleted (rained out)
from the atmosphere to depths below the silicate clouds and to temperatures near 2000 K, far below the 
Gl 229B's visible atmosphere.  As to the suggestion by Griffith \etal, the dependence on wavelength between 0.85 \mic and 
0.93 \mic of the imaginary index of refraction of the hazes inferred by construction was $\sim 5\times$ steeper 
than the corresponding published Tholin or polyacetylene values.  Moreover, though Gl 229B has
in its primary, Gliese 229A, a nearby source of UV photons needed to generate a haze, the late L dwarfs
and the new field methane dwarfs do not, yet they demonstrate similar profound absorption in the blue.

\section{The Pivotal Role of the K I Resonance Doublet at $\sim$7700 \AA}

What motivated Griffith \etal to posit the presence of a photochemical haze was
the need for a continuum absorber to suppress the ``blue" flux and to act shortward
of $\sim$1.0 \mic down to at least 0.8 \mic (Figure 1).  Small Mie scattering particles might quite naturally
have fit the bill.  However, there is another, ready-made, source of continuum opacity, the
red wings of the K I resonance lines at 7665 \AA\ and 7699 \AA, that are in just the 
right place, shortward of the Gl 229B data, with what appear to be just the right strength
and opacity slope, to fully explain the anomalous data.   The K I resonance doublet, due to
the $4s^2S_{1/2}-4p^2P_{1/2,3/2}$ transitions, comes into its own only if other sources
of opacity that dominate in M dwarfs, such as TiO and VO, are absent.  The formation of refractories
and their subsequent rainout accomplishes just that.  

The concept of ``rainout" warrants further explanation: When condensates form, there are two possible 
general categories for their ultimate distribution within an atmosphere.  The condensate
(solid grain or liquid drop) can remain well-mixed with the
atmosphere above the condensation level, or the condensate
can rain out of the atmosphere.  In the latter case, the exact vertical profile
of the condensate depends upon poorly understood microphysical
processes. However, guided by experience with planetary atmospheres,
we expect the condensate to form a cloud layer of some finite
thickness.  Both the condensible gas and its condensate are
thus depleted above the cloud top.  We refer to this
process as ``rainout."  In the other, well-mixed, case the condensate
can continue to react with the gas; we refer to this case as
complete thermochemical equilibrium.  Based on the Gliese 229B data, on experience with 
the planets of our solar system, and on general physical grounds, some 
form of rainout seems to occur.

As shown by Burrows and Sharp (1999), Fegley and Lodders (1996), and Lodders (1999), the alkali metals
are less refractory than Ti, V, Ca, Al, Fe, and Mg and survive in abundance as neutral atoms in substellar
atmospheres to temperatures of 1000 K to 1500 K.  This is below the 1500 K to 2500 K temperature
range in which the silicates, iron, the titanates, corundum, and spinel, etc. condense.
Hence, in the depleted atmospheres of cool brown dwarfs alkali metals quite naturally
come into their own.  Figures 2 and 3 show some representative abundance profiles for the 
%
%
alkali metals Li, Cs, K, and Na, with and without rainout. A \teff=950 K and g=10$^5$ cm s$^{-2}$ Gl 229B atmosphere 
model from Burrows \etal (1997) and Burrows and Sharp (1999) was used.  As demonstrated in 
Figures 2 and 3, with or without rainout elemental potassium and sodium 
persist to low temperatures to near the top of the T dwarf atmosphere.  However, the distributions 
of the atomic alkalis are not the same, with Li and Cs being deeper than Na and K.  At lower temperatures in an atmosphere,
the atomic forms give way to the chlorides and hydroxides ({\it e.g.}, LiOH).   With rainout (Figure 2), below 
$\sim$1000 K both sodium and potassium exist as sulfides (Na$_2$S and K$_2$S) (Lodders 1999).   Without rainout (Figure 3), a situation 
we deem unlikely, complete chemical equilibrium at low temperatures requires that sodium and potassium reside in 
the feldspars, high albite and sanidine.  If such compounds formed and persisted at altitude, then the
nascent alkali metals would be less visible, particularly in T dwarfs.  For either case, as Figures 2 and 3 demonstrate, 
all the elemental alkalis reside in the lower-pressure/lower-temperature reaches of cool, 
substellar atmospheres. 

Lines of Cs I at 8521 \AA\ and 8943 \AA,
Rb I at 7800 \AA\ and 7948 \AA, Li I at 6708 \AA, and Na I at 8183/8195 \AA\ have already been
identified in L dwarf spectra.  In addition, as Figure 4 and Kirkpatrick \etal (1999) demonstrate, the K I doublet is 
%
%
clearly dominant in the spectra of such late L dwarfs as 2MASS-1228, 2MASS-0850,
Denis-0205, and 2MASS-1632.  Therefore, it is quite natural to explain
the steep red slope of the new T dwarfs as being caused by the red wing of a saturated 
and pressure--broadened K I feature.  This is the thesis of our paper.  
Moreover, Gl 229B's $I$ band flux (M$_I\sim$20.76), as measured using WFPC2 
on HST (Golimowski \etal 1998) can be explained by the same
K I resonance feature and the Gl 229B $R$ band flux ($M_R \sim 24.0$), also measured by Golimowski \etal (1998), 
can be explained by a mix of the sodium D lines at 5890 \AA, the lithium line at 6708 \AA,
and the blue tail of the K I resonance doublet.  Note that the 1.25 \mic 
lines of excited K I have also been identified in T dwarfs
(Strauss \etal 1999; Tsvetanov \etal 1999), so that the presence of 
potassium, at least at the higher temperatures in their
atmospheres, is in no doubt.  The alkali metal lines may well prove
to be the key to probing the atmospheric structure of the methane dwarfs,
since their distributions and relative depths ({\it cf.}, Figure 2) will be reflected
in the systematic dependence of T and L dwarf spectra with \teff and $g$.

\section{Profiles of the Alkali Metal Resonance Lines \label{shape}}

Before we proceed with a discussion of our synthetic Gl 229B spectra between 0.3 \mic 
and 1.5 \mic we discuss our algorithm for handling the pressure broadening of the alkali 
metal lines by molecular hydrogen.  The line lists containing 
the wavelength of the transition, the lower excitation energy, the $\log(gf)$ value,
and the quantum numbers of the participating states were obtained from the 
Vienna Atomic Line Data Base (Piskunov \etal 1995).  The major transitions of immediate 
relevance are those that correspond to the Na D lines at 5890 \AA\ and the K I resonance
lines at 7700 \AA.  Given the high H$_2$ densities in brown dwarf 
atmospheres, the natural widths (for Na D, $\sim$0.12 m\AA) and 
Doppler widths of these lines are completely overwhelmed by collisional broadening.  However,
in general the line shapes are determined by the radial dependence of the difference of the perturber/atom potentials for
the lower and upper atomic states (Griem 1964; Breene 1957,1981) 
and these are rarely known.  The line cores are determined by distant 
encounters and are frequently handled by assuming a van der Waals interaction 
potential with an adiabatic impact theory (Weisskopf 1933; Ch'en and Takeo 1957; 
Dimitrijevi\'{c} and Peach 1990) and the line wings are determined by 
close encounters and are frequently handled with a statistical theory (Holtzmark 1925; Holstein 1950).
The transition between the two regimes is near the frequency shift ($\Delta\nu$), or detuning, associated with
the perturbation at the so-called Weisskopf radius ($\rho_{\rm w}$), from which the collision cross section employed in
the impact theory is derived (Spitzer 1940; Anderson 1950).  In the simple impact theory, the line core
is Lorentzian, with a half width determined by the effective collision frequency, itself  
the product of the perturber density, the average relative velocity of the atom and the perturber ($v$), and 
the collision cross section ($\pi\rho_{\rm w}^2$).  If the frequency shift ($\Delta\nu$) due to a single perturber
is given by $C_n/r^n$, where $r$ is the interparticle distance, then $\rho_{\rm w}$ is determined from the condition 
that the adiabatic phase shift, $\int^{\infty}_{-\infty}2\pi\Delta\nu  dt$, along a classical straight--line trajectory, with
an impact parameter $\rho_{\rm w}$, is of order unity.  This yields $\rho_{\rm w}\propto (C_n/v)^{1/(n-1)}$.
For a van der Waals force, $n=6$.  In the statistical theory, the line shape is a power law that
goes like $1/\Delta\nu^{\frac{n+3}{n}}$, and this is truncated (cut off) by an exponential 
Boltzmann factor, $e^{-V_0(r_s)/kT}$, where $V_0(r_s)$ is the ground-state perturbation
at the given detuning.  The detuning at the transition between the impact and
statistical regimes is proportional to $(v^6/C_n)^{0.2}$ (Holstein 1950).  

All this would be academic, were it not that for the Na/H$_2$ pair the simple theory in the core 
and on the red wing is a good approximation (Nefedov, Sinel'shchikov, and Usachev 1999).  
We use this theory here.  For the Na D lines perturbed by H$_2$, we obtain from Nefedov, 
Sinel'shchikov, and Usachev (1999) a $C_6$ of $2.05\times10^{-32}$ $cgs$ and a transition
detuning, in inverse centimeters, of 30 cm$^{-1}(T/500 K)^{0.6}$, where $T$ is the 
temperature.  For the K I resonance lines, we scale from the Na D line data, using
a $C_6$ of $1.16\times10^{-31}$ {\it cgs}, itself obtained from the theory of Uns\"old (1955). 
This procedure yields a transition detuning for the 7700 \AA\ doublet of 20 cm$^{-1}(T/500 K)^{0.6}$.
From Nefedov, Sinel'shchikov, and Usachev (1999), we see that for a variety of perturbing
gases the exponential cutoff for the Na D lines can be (for temperatures of 1000--2000 K) 
a few $\times 10^{3}$ cm$^{-1}$.  The difference between 5890 \AA\ and 
7700 \AA, in inverse wavenumbers, is $\sim$4000 cm$^{-1}$ 
and that between 7700 \AA\ and 1.0 \mic is only 3000 cm$^{-1}$.  Hence, it is reasonable to
expect that the detunings at which the line profiles are cut off can be much larger
than the Lorentzian widths or the impact/statistical transition detunings of tens of cm$^{-1}$.  
Since we as yet have no good formula for the exponential cutoff term, we assume that it is
of the form $e^{-qh\Delta\nu/kT}$, where $q$ is an unknown parameter.
Comparing with the examples in Nefedov, Sinel'shchikov, and Usachev (1999), $q$ may be
of order 0.3 to 1.0 for the Na/H$_2$ pair.  Without further information or guidance,
we assume that it is similar for the K/H$_2$ pair.  We stress that this algorithm is
merely an ansatz and that a more comprehensive theory based on the true perturber potentials 
is sorely needed. Nevertheless, whatever the detailed line shape, as we will show, the basic
conclusion that the K I resonance doublet is the ``mystery" absorber in the near-infrared
spectra of T dwarfs seems robust.

%
%
Figure 5 depicts our opacity spectrum versus wavelength for the K I doublet at $T=2000$ K and 1 bar
pressure, for three different parameterizations.  Included are other, non-resonant, potassium lines
excited by the high temperature.  The solid line is for the standard Lorentzian theory, without
a transition to the $3/2$-power law.  The other lines depict the line profile, corrected 
in the wings and with an exponential cutoff, using either $q=1.0$ or $q=0.5$.  Note that 
for $q=1.0$, beyond the core the transition to the $3/2$-power law from the Lorentzian ``$2$"-power law
results in a larger opacity, which is then brought below the Lorentzian profile by the
exponential.  The same behavior is seen for $q=0.5$, but the exponential brings the profile
below the Lorentzian at a larger detuning.   In all cases, the slope redward of the 7700 \AA\
doublet between 0.8 \mic and 1.0 \mic is of just the sort to affect the desired reddening.

\section{Model Spectra for Gliese 229B}

Gliese 229B has been well--studied from the red to $\sim$10 \mic and has a well--measured distance (5.8 parsecs).
Hence, with moderate resolution spectra in hand and absolute fluxes it is natural to use Gl 229B to demonstrate our 
basic thesis about the near-infrared and optical spectra of the entire T dwarf class.
On Figure 1, in addition to depicting the ``Clear" atmosphere fluxes from 0.5 \mic to $\sim$1.45 \mic,
we give the synthetic spectra for four different models of Gl 229B that include the lines of the alkali metal atoms,
in particular the K I doublet at 7700 \AA, employing the formalism of \S\ 3.  The Gl 229B data are taken from Leggett \etal (1999).
Rainout abundance profiles of the alkali metals were used ({\it cf.}, Figure 2).
As Figure 1 demonstrates, the fit between 0.85 \mic and 1.0 \mic is quite good and does not seem to require a layer of red particulates.
We were guided in our choices of \teffs, gravities, and metallicities for these models by the work of Marley \etal (1996),
Allard \etal (1996), and Griffith \etal (1998).  Three of the models, those with higher fluxes around 0.7 \mic and lower fluxes
near 1.0 \mic used a $q$ of 0.4 and assumed alkali metal abundances of 0.3 times 
solar (Anders and Grevesse 1989).  They had \teff/$g$ pairs
of [900 K/10$^5$ ($cgs$)], [950 K/10$^5$ ($cgs$)], and [800 K/$3\times 10^4$ ($cgs$)].  From Burrows \etal (1997), the 
ages and masses for these models are 1.46 Gyr/34.9 \mj, 1.26 Gyr/35.3 \mj, and 0.44 Gyr/14.8 \mj, respectively.
The fourth model, that with the lowest fluxes near 0.7 \mic and the highest fluxes near 1.0 \mic, employed a
$q$ of 1.0, had a \teff/$g$ pair of [950 K/10$^5$ ($cgs$)], and assumed that the 
alkali metal abundances were solar.  Note that values for q of 0.4 and 1.0 are merely possibilities and that we do
not necessarily advocate either one.  As stressed in \S 3,   
the far wings of the neutral alkali metal lines, perturbed by molecular 
hydrogen, have not been properly calculated.  In fact, to our knowledge there is no
other environment in which the neutral alkali line strengths at 1000-3000 \AA\
detunings have ever before been needed.

Figure 6 depicts the behavior of the theoretical spectra 
%
%
as a function of $q$, from 0 to 1, for the [950 K/10$^5$ ($cgs$)] model (with 0.3$\times$solar alkali 
metal abundances).  Also included is the predicted spectrum for the uncorrected 
Lorentzian K I line profile (dotted).  Though Figure 6 depicts a range of near-infrared 
and optical spectra, the character of these spectra is similar.  However, as a consequence of the current 
ambiguity in the proper shape of the 7700 \AA\ profile, which we have parameterized with $q$,  
we can not yet tie down the alkali metal abundances to better than perhaps a factor of three.  
Nevertheless, the good fits indicated in Figure 1 are quite compelling.  Rather stunningly, 
both the measured WFPC2 $I$ band flux near 0.814 \mic and the WFPC2 $R$ flux near 0.675 \mic 
(Golimowski \etal 1998) fit as well.  Hence, we conclude that the $I$ band flux is 
determined by the red wing of the K I resonance doublet and that the $R$ band is
a consequence of the red wings of the Na D lines ($\sim 5890$ \AA), the blue wings of the K I doublet,
and, perhaps, the lithium line at 6708 \AA.   The broad width of the $I$ band filter extends it
to beyond the saturated center of the doublet at 7700 \AA.  Our theoretical models have absolute $I$ band magnitudes from 21.0 
to 21.3, compared with the measured value of $\sim$20.76.  Given the calibration problems when an underlying spectral slope
is so steep and so unlike that of the standard stars, we conclude that the fit is indeed good.
As Figure 1 demonstrates, our $R$ band predictions bracket the $R$ band measurement.

We emphasize that one should be very careful about interpreting band magnitudes
as fluxes.  There are two points to make in this
regard: 1) bands are spread over a sometimes broad range of wavelengths
and, 2) magnitudes are calibrated on stars with certain underlying
spectra that may not be similar to the spectrum of the object under
study.  If these spectra are very different, large errors
can be made in assigning magnitudes, as well as in assigning
fluxes at a given wavelength in the band.  A quick look at the
spectrum of Gl 229B from 0.8 to 1.0 $\mu$m will show why this
should be of concern; the calibration stars can not have such
steep slopes and the errors in the measured magnitudes might be very large,
many tenths of magnitudes.  This, and extravagant band width, 
are particularly relevant for the Gunn $r$ and $i$
band measurements of Matthews \etal (1996).  Given this,
we compare with just the WFPC2 $R$ and $I$ band data of Golimowski \etal  
as a set from a single group and depict the
breadth of these bands on our figures.

We note in passing that the discrepancy between the depth of the predicted and measured flux trough 
between the $Z$ ($\sim$1.05 \mic) and the $J$ ($\sim$1.25 \mic) bands may be resolved 
with a lower abundance of water and, hence, oxygen (Griffith \etal 1998)
and that the shape of the $J$ band can not be fit without the methane bands at $\sim\!1.15-1.2$ \mic and
$\sim\!1.3$ \mic to sculpt it.  Hence, the shape of the $J$ band in T dwarfs is not just a 
consequence of bracketing by H$_2$O absorption features, but requires methane and is another signature of methane
is their atmospheres.  However, this point, as well as the inferred elemental abundances, are not the subject of the 
current paper and we defer a fuller discussion to a later work.  

%
%

The variety of models that fit the Gl 229B data, given the current state of the theory
and observations, was noted in Marley \etal (1996) and Allard \etal (1996) and reflects
the fact that different models with different \teff/$g$ pairs can have similar
atmospheric temperature--pressure profiles and luminosities.  The atmospheres of substellar objects
are convective at depth and radiative on the periphery.  
For a given composition, brown dwarfs and extrasolar giant planets (EGPs; Burrows \etal 1995)
are a two-parameter family; given two independent quantities such as \teff, luminosity, radius,
gravity, or age, one can derive, with theory, any of the others.   In particular, as 
discussed in Hubbard (1977) and Saumon \etal (1996), in an approximate sense, the entropy ($S$) 
at depth is a function of \teff and $g$ and this entropy determines the temperature/pressure
profile.  (This is not to say that the atmospheres are adiabatic, merely that the T/P profile 
of the outer radiative zone can be calculated from $S$, \teff, and $g$.)  Hubbard (1977) showed
that $S$ is approximately a function of the combination $T_{\rm eff}^{0.95}/g^{1/6}$.  From Marley
\etal (1996), we obtain rough power-law relations for mass ($M$), radius ($R$), and age ($t$), 
as a function of $g$ and \teff :
\begin{eqnarray}
M &=& 35 M_{\rm J}\Bigl(\frac{g}{10^5}\Bigr)^{0.64}\Bigl(\frac{T_{\rm eff}}{10^3}\Bigr)^{0.23}
\nonumber \\
R &=& 6.7\times10^4{\rm km}\Bigl(\frac{10^5}{g}\Bigr)^{0.18}\Bigl(\frac{T_{\rm eff}}{10^3}\Bigr)^{0.11}
\nonumber \\
t &=& 1.1 {\rm Gyr}\Bigl(\frac{g}{10^5}\Bigr)^{1.7}\Bigl(\frac{10^3}{T_{\rm eff}}\Bigr)^{2.8}\, ,
\end{eqnarray}
where $g$ is in cm s$^{-2}$. From these equations, and the dependence of $S$ on \teff and $g$
cited above, we derive approximately that, for a given core entropy and, hence, for a given T/P profile,
$M \sim t^{0.55}$.  Using the same set of power laws, we can derive that, for a given
luminosity, $M \sim t^{0.43}$.  Hence, since these two power-law relations are similar,
fits to Gl 229B roughly define a trajectory in $M$-$t$ space with a $\sim0.5$ power. 
This is why low-mass/short-age models and high-mass/long-age models both fit Gl 229B,
at least for the low--resolution comparisons thus far attempted.  In principle, this degeneracy can 
be broken once the theory and the observations are better constrained.  Figure 7
%
%
shows temperature/pressure profiles for a few models from Burrows \etal (1997).
The higher curves are for higher \teffs and lower gravities.
Note that the lower-\teff, lower-$g$ model [800 K/$3\times 10^4$ ($cgs$)] 
is in the same vicinity as the higher-\teff, higher-$g$ model [900 K/10$^5$ ($cgs$)],
both of which are good fits to Gl 229 B (Figure 1).  Note also that the  [1100 K/$5\times 10^4$ ($cgs$)], 
[400 K/$5\times 10^4$ ($cgs$)], and [800 K/$10^5$ ($cgs$)] models depicted for comparison do not fit Gl 229B, 
as we might have surmised from Figure 7.

\section{The Sloan Dwarf: SDSS 1624+00}

The models and data depicted in Figure 1 capture the Cs lines at 8521 \AA\ and 8943 \AA, indicating that 
cesium is in Gl 229 B's atmosphere in modest abundance between 1200 K and 1400 K.  
The depths to which cesium can be found are shown in detail for a specific model in Figures 2 and 3   
and approximately for the variety of models by the position of the intercept in Figure 7 of the ``Cs/CsCl=1" line
with a given T/P profile.  Similarly, the depths of atomic K, Na, Rb, and Li can be found.  Shown on Figure 7
along with the Cs/CsCl=1 line is a corresponding complete equilibrium (no rainout) ``K/KCl=1" trajectory.
As Figure 2 demonstrates, with rainout (a more realistic situation), atomic potassium  persists to 
much lower temperatures, perhaps shifting the K/KCl=1 line on Figure 7 down by 200--300 K.
Nevertheless, qualitatively and importantly, the lower the T/P profile in Figure 7, the deeper are the
atomic cesium and potassium.  Depth and higher pressure imply a higher column density
of H$_2$O, CH$_4$, and H$_2$ and, hence, a higher optical depth of these competing absorbers.  
Therefore, for a given composition, at lower \teff and higher $g$, both potassium and cesium are
more deeply buried than at higher \teff and lower $g$.  The expected upshot is the gradual
diminution with decreasing core entropy of the strengths of the cesium and potassium features.  A decrease in the
effect of the K I resonance doublet at 7700 \AA\ would result in an increase in the influence below
1.0 \mic of the H$_2$O, CH$_4$, and H$_2$ features that are not as steep as the red wing of the doublet.
As a result, the spectrum short of $\sim$0.95 \mic would shallow and the 7700 \AA\ region of the 
spectrum would fill in.  The Rb lines at 7800 \AA\ and 7948 \AA\ may make an 
appearance in the K I trough, but they will still be rather weak.
Note that, for a given \teff/$g$ pair, lowering the abundances of the alkali metals would raise the flux in the 0.7-0.95 \mic
region, but it might also make the $Z-J$ color bluer than observed for the current crop of T dwarfs.  
Furthermore, the presence of the potassium line at 1.25 \mic both in Gl 229B and in the two Sloan
dwarfs makes this explanation suboptimal.   Nevertheless, potential variations in abundances and 
metallicity from object to object do complicate the analysis.

The shallowing of the spectrum shortward of 0.9 \mic and the weakening 
of the cesium features at 8521 \AA\ and 8943 \AA\ are indeed seen in the preliminary spectrum of SDSS 1624+00 
(Strauss \etal 1999).  Strauss \etal note that the spectra of Gl 229B and SDSS 1624+00 seem uncannily
similar and conclude that they must have similar mass, age, and \teff.  Given the lower apparent flux of
SDSS 1624+00, they put this object further away than Gl 229B, at $\sim$10 parsecs.  However, theory implies 
the $J-H$ and $J-K$ colors don't change hugely in the 600 K to 1200 K range (Burrows \etal 1997).
As a consequence, for a given gravity or mass, a decrease in \teff merely, though approximately, translates the absolute
spectrum down to lower fluxes.  Hence, there is a need for more subtle diagnostics of a substellar object's
properties.  Such diagnostics may be the strengths of the alkali metal lines in the optical and near infrared. 
Figure 8 compares the observed SDSS 1624+00 spectrum with a theoretical model spectrum, as well 
as with the observed Gl 229B spectrum. The model (dotted) has an atmosphere with \teff=700 K and 
$g$=$3\times10^4$ cm s$^{-2}$. Its core entropy is below those of the Gl 229B models.
We theorize that the shallower near-infrared spectrum 
of SDSS 1624+00 with respect to Gl 229B, as well as the weakness of the Cs lines,
are both consequences of a lower core entropy T/P profile.   This could mean a lower \teff, a higher $g$ or
mass, or some suitable combination that buries the tops of the potassium and cesium distributions at higher 
pressure and, hence, larger H$_2$O optical depths than found in Gl 229B.  However, given the  
remaining theoretical ambiguities, we can't completely rule out the possibility that 
lower K and Cs abundances are the explanation.  Note that it is still possible that \teff for SDSS 1624+00 is
much larger than 700 K. However, given its provisional cesium line 
strengths, we would then require its gravity, and presumably its mass, to
be correspondingly higher to achieve the lower core entropy we suggest can explain its spectrum. 
Importantly, if improved spectra were to show that Cs is
much stronger in SDSS 1624+00 than its preliminary, low--resolution spectrum now indicates,
a higher \teff, perhaps higher than that of Gl 229B, would be indicated.   This emphasizes the importance 
of the alkali metal lines as diagnostics of the T dwarfs and the need to obtain better spectra.
Note that with the [700 K/$3\times10^4${\it cgs}] model, the absolute $Z$ and $J$ band fluxes of 
the SDSS 1624+00 are below those for Gl 229B and match only near $\lambda=0.85$ \mic.  If the \teff of
SDSS 1624+00 is indeed lower than that of Gl229B, SDSS 1624+00 could be closer than the 5.8 parsecs
of Gl 229B.  A smaller distance would increase the inferred number density of SDSS 1624+00-like objects,
perhaps beyond what is credible.
Hence, obtaining a parallax for this new T dwarf will be a crucial prerequisite for 
future substantial improvements in characterizing it. 

Importantly, any \teff estimate for SDSS 1624+00 must be confirmed by a comparison
of models and observations at longer wavelengths, similar to that
performed for Gl 229B (Marley {\it et al.} 1996; Allard {\it et al.} 1996).  As with Gliese 229B,
simultaneously achieving a satisfactory fit to the optical flux, the
near-infrared windows, and absorption features will be challenging.
In particular, if SDS 1624+00 has a lower \teff than Gliese 229B,
the water and methane band depths will likely be deeper.  Yet
the methane opacity database at 1.7 \mic is too poor to permit detailed
comparisons and the 2.3 \mic band is also influenced by H$_2$ opacity, the
relative strength of which depends upon the metallicity.  L$^{\prime}$
band spectra, which probe the methane fundamental band at 3.3 \mic
where there are no competing absorptions and the opacity database
is better understood, may provide the best point of comparison.

However, for a given mass and composition, the ``history" of the alkali metal lines in L and T dwarfs may follow
the following sequence:  At higher \teffs (in the L dwarf range), the depletion of the refractory elements allows the strengths
of the alkali metal lines to wax.  The cesium lines peak before reaching the \teff of Gl 229B, though they are still
prevalent in such atmospheres, and potassium determines the character of the spectrum 
from 0.7 \mic through 1.0 \mic.  As \teff decreases further, the alkali metals are 
buried progressively more deeply and begin to wane in strength.   Due to this burial of potassium,
the slope of the spectrum from 0.8 \mic to 1.0 \mic decreases.  Since the abundance distribution
of the atomic form of each alkali metal is different (Figures 2 and 3), the \teffs at which the lines of a given  
alkali start to wane will be different, with the effects of Li, Cs, Rb, K, and Na waning in approximately that
order with decreasing \teff (at a given mass and metallicity).  The specific numbers will depend upon 
the actual effects of rainout on the alkali metal profiles, but the systematics should not.  
Eventually, perhaps below \teffs of 500-600 K, 
the effects of the neutral alkali metals are eclipsed, with clouds of H$_2$O eventually 
effecting the next important change in the character of brown dwarf atmospheres.  
Given this systematics, whatever the true \teff of SDSS 1624+00, we would predict 
that both above Gl 229 B's \teff, near the \teff juncture between the L dwarfs and the T dwarfs, and below 
a \teff of $\sim$800 K, the troughs around the K I resonance doublet at 7700 \AA\ would be partially filled
and the slope of the ``optical'' spectra between 0.8 and 1.0 \AA\ would shallow.    

\section{Summary}

We conclude that the anomalous absorption seen in the near-infrared and optical spectra of all the T dwarfs discovered to date is 
due to the red wings of the saturated K I resonance lines at 7700 \AA.  This theory also explains 
the WFPC2 $I$ and $R$ flux measurements made of Gl 229B, with the Na D lines at 5890 \AA\ helping to determine the strength of
the $R$ band.  There are still ambiguities in the \teffs, gravities,
and compositions of Gl 229 B, in particular, and of T dwarfs, in general, but a sequence
from Gl 229B to SDSS 1624+00  of decreasing core entropy is seen to be consistent with the expected systematics
of the temperature/pressure and alkali--metal abundance profiles.  (However, lower K and Cs abundances for
the Sloan dwarf or a reassessment of its cesium line strengths once higher--resolution spectra become available
can not yet be ruled out.) Silicate grains are expected at depths of 1500 K to
2500 K.  Their formation is inferred in the late M dwarfs (Jones and Tsuji 1997)
and their eventual burial is seen in the appearance at lower \teffs of the new 
L spectral class (Kirkpatrick \etal 1999).  Complete refractory element depletion  
ushers in a phase of neutral alkali metal dominance at short wavelengths, which persists until the tops of the
alkali metal distributions are buried at higher pressures and higher H$_2$O, CH$_4$, and H$_2$ optical depths.
This occurs in low--entropy atmospheres, though low alkali metal abundances can mimic the same effect somewhat.
We note that if the Lorentzian theory for the line shapes of the Na D lines were used, due to the 
high relative abundance of sodium, their influence would stretch 7000 cm$^{-1}$ to 1.0 \mic and would flatten the top
of the $Z$ band, contrary to observation.  This is indirect evidence for the action of an ``exponential cutoff"
on the Na D line shape.  Even with such a cutoff, the K I resonance line dominates the spectra of
Gl 229 B and its ilk from 0.7 \mic to 1.0 \mic.   For Gl 229 B,  we predict the presence of a 
saturated absorption line around 7700 \AA, with higher fluxes on either side.  We also expect a deep, broad trough
around the Na D lines at 5890 \AA\ (see Figure 1).  

At the time of this writing, we became aware of a paper by Tsuji \etal (1999) that has also concluded that the K I
feature is the major, ``mystery" absorber at short wavelengths.  We wholeheartedly concur with this conclusion.
However, contrary to Tsuji \etal, we determine 
that dust may not be required to achieve a good fit shortward of 1.0 \mic.   Moreover, 
we believe that the Gliese 229B $I$ band flux is well--fit by the red wing of the K I doublet alone.
In addition, if the Na D lines are included in a natural way, an acceptable 
fit to the WFPC2 $R$ band flux is achieved as a byproduct.  However, filling in the troughs between the $Z$ and
$J$ bands and between the $H$ and $K$ bands may still require some combination of grains, as Tsuji \etal suggest, 
and sub--solar metallicity, though sub--solar metallicity and improvements in the molecular
opacity databases alone may be sufficient.  If silicate dust is important, it could be because of
upwelling from below, a very non-equilibrium process that is not
easily modeled.  In addition, optical constants, particle shapes,
and particle sizes would need to be known.  There is no credible
guidance concerning the particle size; the guessed average particle
radius could off by one or two orders of magnitude.  This ambiguity
translates into a correspondingly large ambiguity in the dust opacity.
As Griffith \etal (1998) demonstrate, a sufficiently arbitrary dust model
can be found to fit the Gl 229B spectrum. 
Nevertheless, a possible model might be one in which a
small population of dust from minor species at low pressures warms
the atmosphere near 1 bar, thus decreasing the depths of the water and methane
absorption bands which originate near this pressure, yet providing little opacity
elsewhere.  However, we reiterate that, given the new-found importance of the 
K I feature, there is much less reason to evoke a population of dust or grains 
to fit methane dwarf spectra.

Note that in this paper we have considered only objects in isolation.  However, we expect that those EGPs that achieve  
\teffs in the 800--1300 K range due to stellar insolation will also show signatures of the                           
neutral alkali metals.   Recently, Charbonneau \etal (1999) have put an upper limit of 0.3 to the geometric
albedo at $\lambda=0.48$ \mic of the planet orbiting $\tau$ Boo.  This is below 
some published predictions (Marley \etal 1999), though not others (Seager and Sasselov 1998),
and may indicate the presence in its atmosphere of sodium and absorption by the Na D lines (Sudarsky, Burrows, and Pinto 1999).
Our preliminary estimate of the geometric albedo at $\lambda=0.48$ \mic of such a planet, due to
the influence of the neutral alkali metals, is $< 0.1$.
Since stellar insolation is bound to create hazes (Marley 1998), 
such as absorb in the blue and UV in Jupiter, detailed reflection spectra
of $\tau$ Boo and similar ``close--in" EGPs will be needed to disentangle the 
relative contribution of the alkali metals to EGP albedos. 

\acknowledgements

We thank Richard Freedman, Phil Pinto, Dick Tipping, Jonathan Lunine, Bill
Hubbard, David Sudarsky, Didier Saumon, Caitlin Griffith, Ben Oppenheimer, Sandy Leggett, 
and Jim Liebert for many useful contributions, as well as for good advice.
This work was supported in part by NASA grants NAG5-7499, NAG5-7073, NAG2-6007,
and NAG5-4987, as well as by an NSF CAREER grant (AST-9624878) to M.S.M.

\clearpage

\begin{figure}
\epsscale{1.05}
\plotone{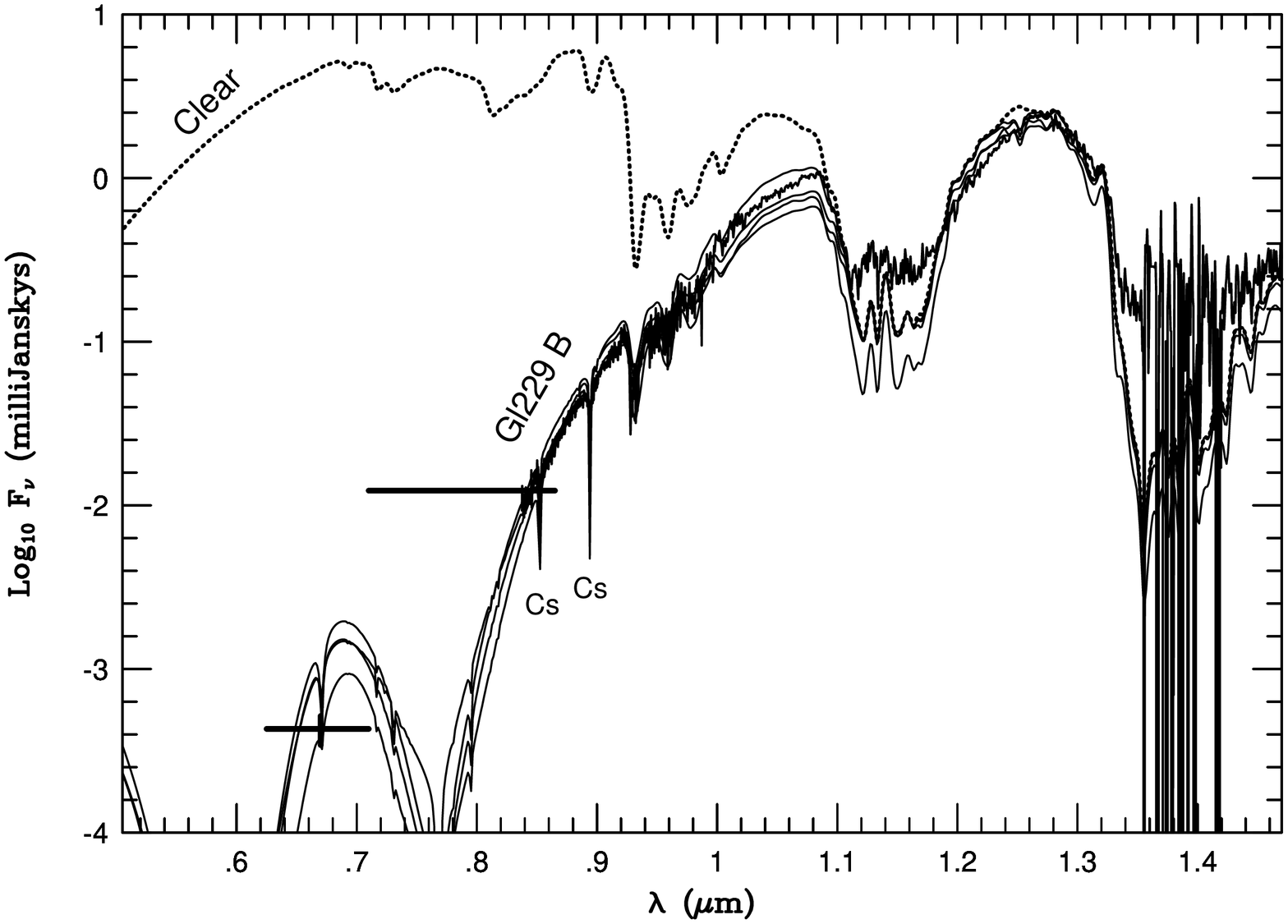}
\caption{
The log of the absolute flux (F$_\nu$) in milliJanskys versus wavelength 
($\lambda$) in microns from 0.5 \mic to 1.45 \mic for Gliese 229 B,
according to Leggett \etal (1999) (heavy solid), and for four theoretical models (light solid) described in the text.  Also
included is a model, denoted ``Clear" (dotted),  without alkali metals and without any ad hoc absorber
due to grains or haze.  The horizontal bars near 0.7 \mic and 0.8 \mic denote the 
WFPC2 $R$ and $I$ band measurements of Golimowski \etal (1998).
\label{fig:1}}
\end{figure}

\begin{figure}
\epsscale{0.7}
\plotfiddle{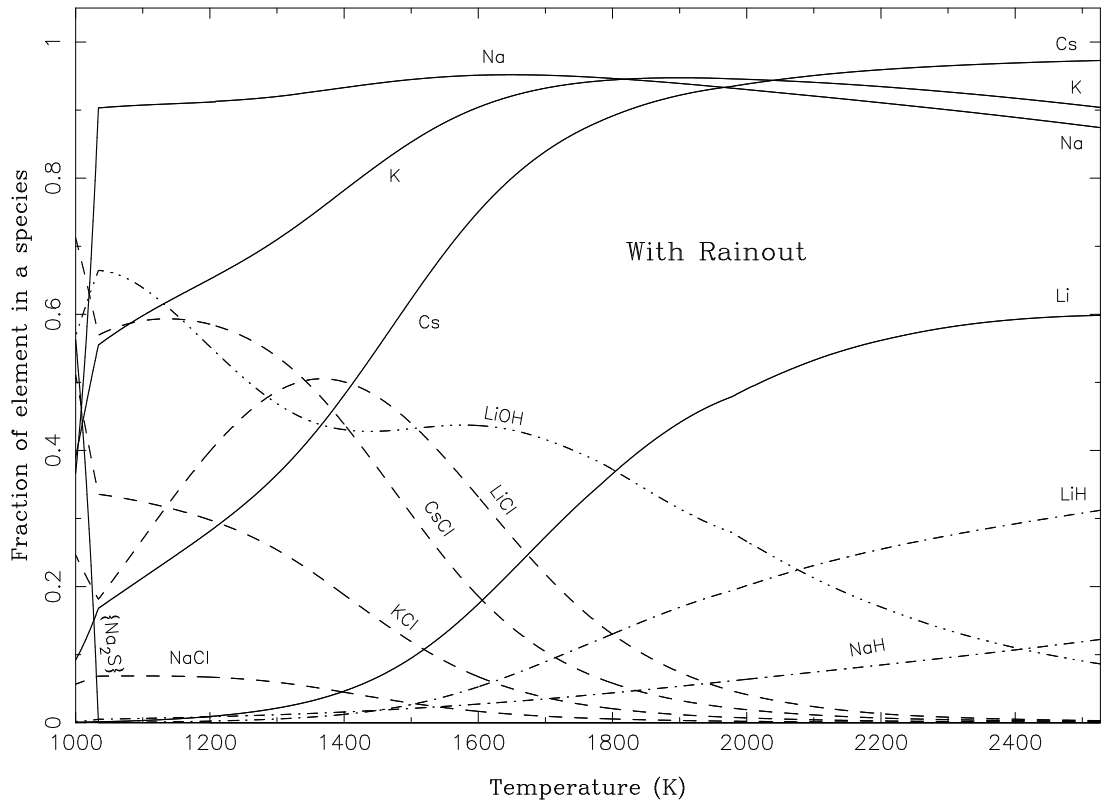}{4.0in}{-90}{60}{60}{-230}{350}
\caption{
The fractional abundances of different chemical species involving
the alkali elements Li, Na, K and Cs for a Gliese 229B model, with rainout
as described in Burrows and Sharp (1999).  
The temperature/pressure profile for a \teff=950 K and
$g=10^5$ cm s$^{-2}$ model, taken from Burrows \etal (1997), was used.
Each curve shows the fraction of the alkali element in the indicated form 
out of all species containing that element.
All species are in the gas phase except for the condensates, which are in braces \{
and \}.  The solid curves indicate the monatomic gaseous species Li, Na,
K and Cs, the dashed curves indicate the chlorides,
the dot-dashed curves indicate the hydrides and the triple dot-dashed curve
indicates LiOH.
Due to rainout, at lower temperatures there is a dramatic difference
with the no--rainout, complete equilibrium calculation (Figure 3); high albite
and sanidine do not appear, but instead at a much lower temperature the
condensate Na$_2$S (disodium monosulfide) forms, as indicated by the solid
line in the lower left of the figure.  The potassium equivalent, K$_2$S, also
forms, but it does so below 1000 K and is not indicated here.   
The difference between this Figure and Figure 3 is that almost all the silicon
and aluminum have been rained out at higher temperatures, so that no
high albite and sanidine form at lower temperatures.
\label{fig:2}}
\end{figure}

\begin{figure}
\epsscale{0.7}
\plotfiddle{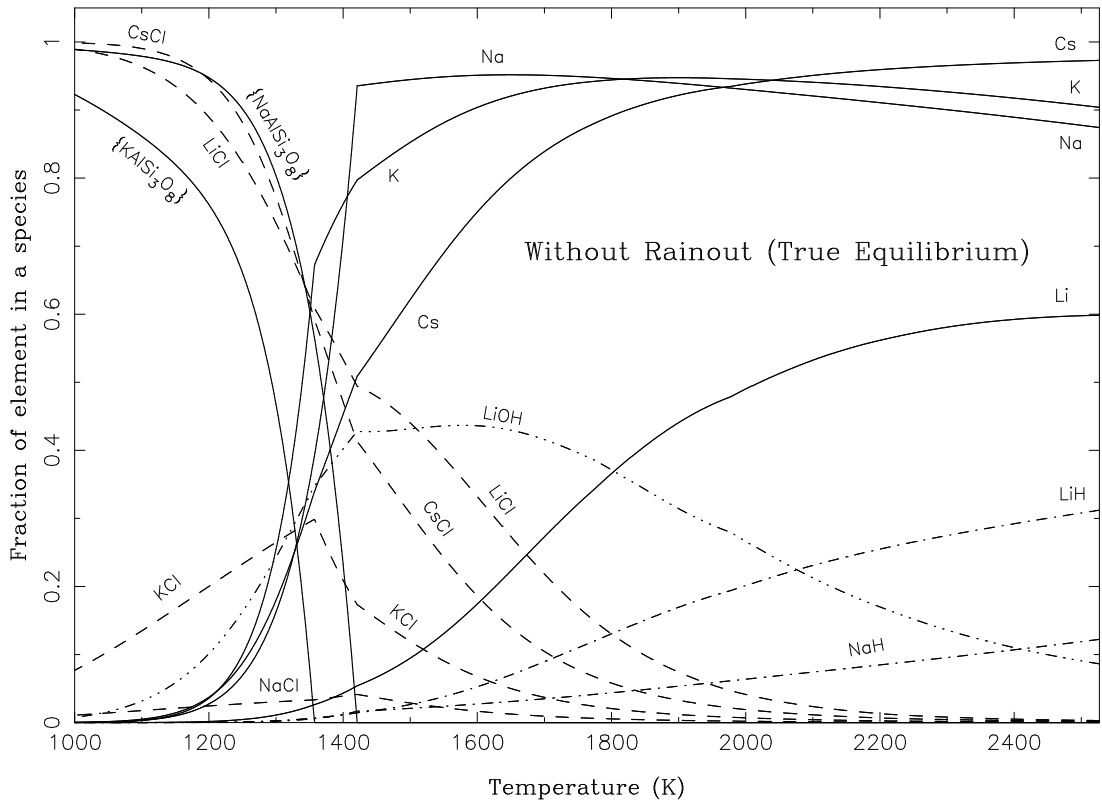}{4.0in}{-90}{60}{60}{-230}{350}
\caption{
The fractional abundances of different chemical species involving
the alkali elements Li, Na, K and Cs for a Gliese 229B model, assuming complete (true)
chemical equlibrium and no rainout (disfavored).  The temperature/pressure profile for a \teff=950 K and
$g=10^5$ cm s$^{-2}$ model, taken from Burrows \etal (1997), was used.
Each curve shows the fraction of the alkali element in the indicated form
out of all species containing that element, {\it e.g.}, in the case of sodium, the
curves labeled as Na, NaCl, NaH and NaAlSi$_3$O$_8$ are the fractions of that
element in the form of the monatomic gas and three of its compounds.  All
species are in the gas phase except for the condensates, which are in braces \{
and \}.  The solid curves indicate the monatomic gaseous species Li, Na,
K and Cs and the two condensates NaAlSi$_3$O$_8$ and KAlSi$_3$O$_8$, {\it i.e.}, high
albite and sanidine, respectively, the dashed curves indicate the chlorides,
the dot-dashed curves indicate the hydrides and the triple dot-dashed curve
indicates LiOH.
\label{fig:3}}
\end{figure}

\begin{figure}
\epsscale{1.05}
\plotone{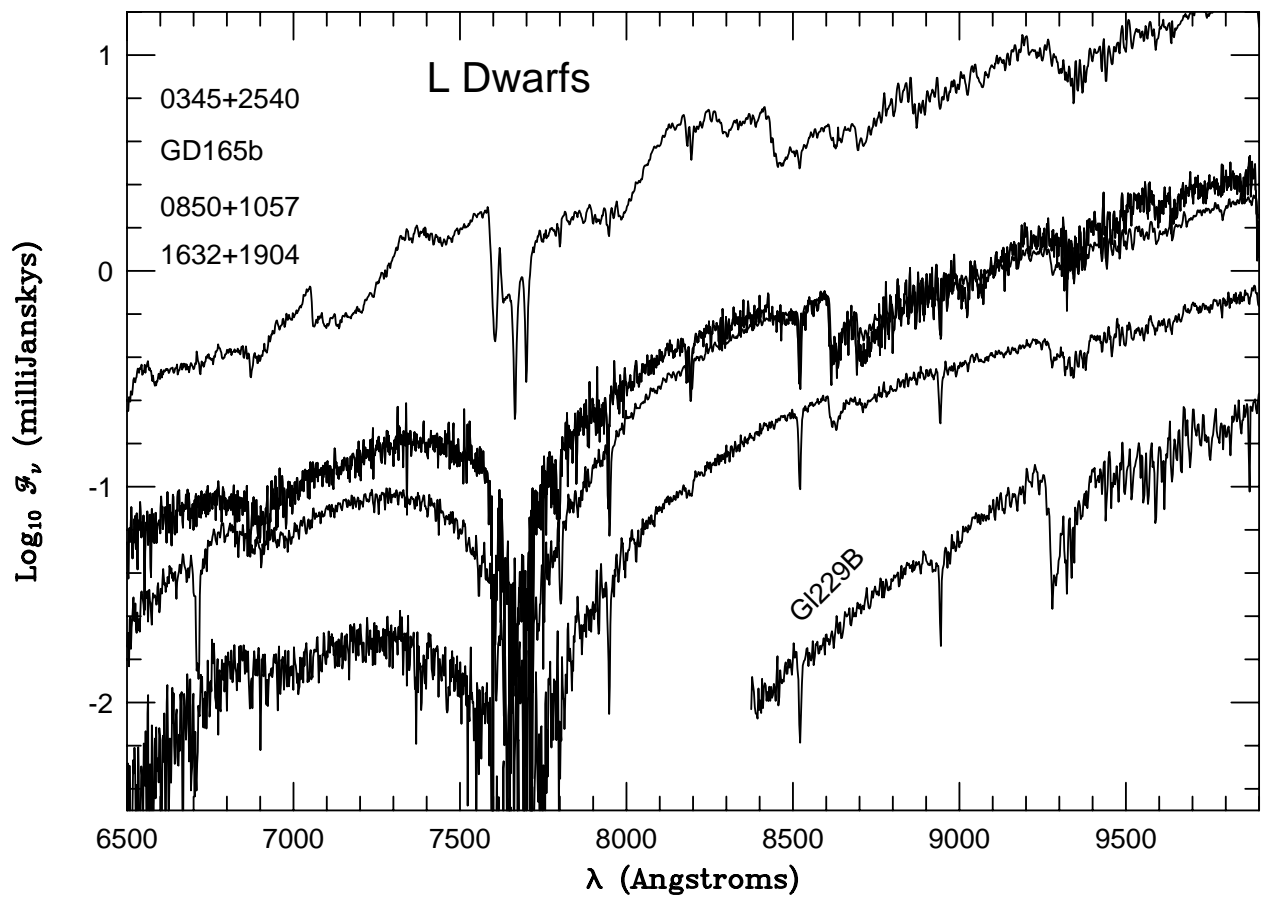}
\caption{
The log of the absolute flux (F$_{\nu}$) in milliJanskys versus wavelength 
($\lambda$) in \AA\ from 6500 \AA\ to 10000 \AA\ for Gliese 229 B
and representative L dwarfs from Kirkpatrick \etal (1999).  Included are 2MASP-J0345 (L0),
GD 165b (L4), 2MASSs-J0850 (L6), and 2MASSW-J1632 (L8), in order of decreasing general flux level.  
The L dwarf spectra were put on an absolute scale using
the distances and apparent $I$ band magnitudes in Kirkpatrick \etal and should be used with extreme caution.
\label{fig:4}}
\end{figure}

\begin{figure}
\epsscale{1.05}
\plotone{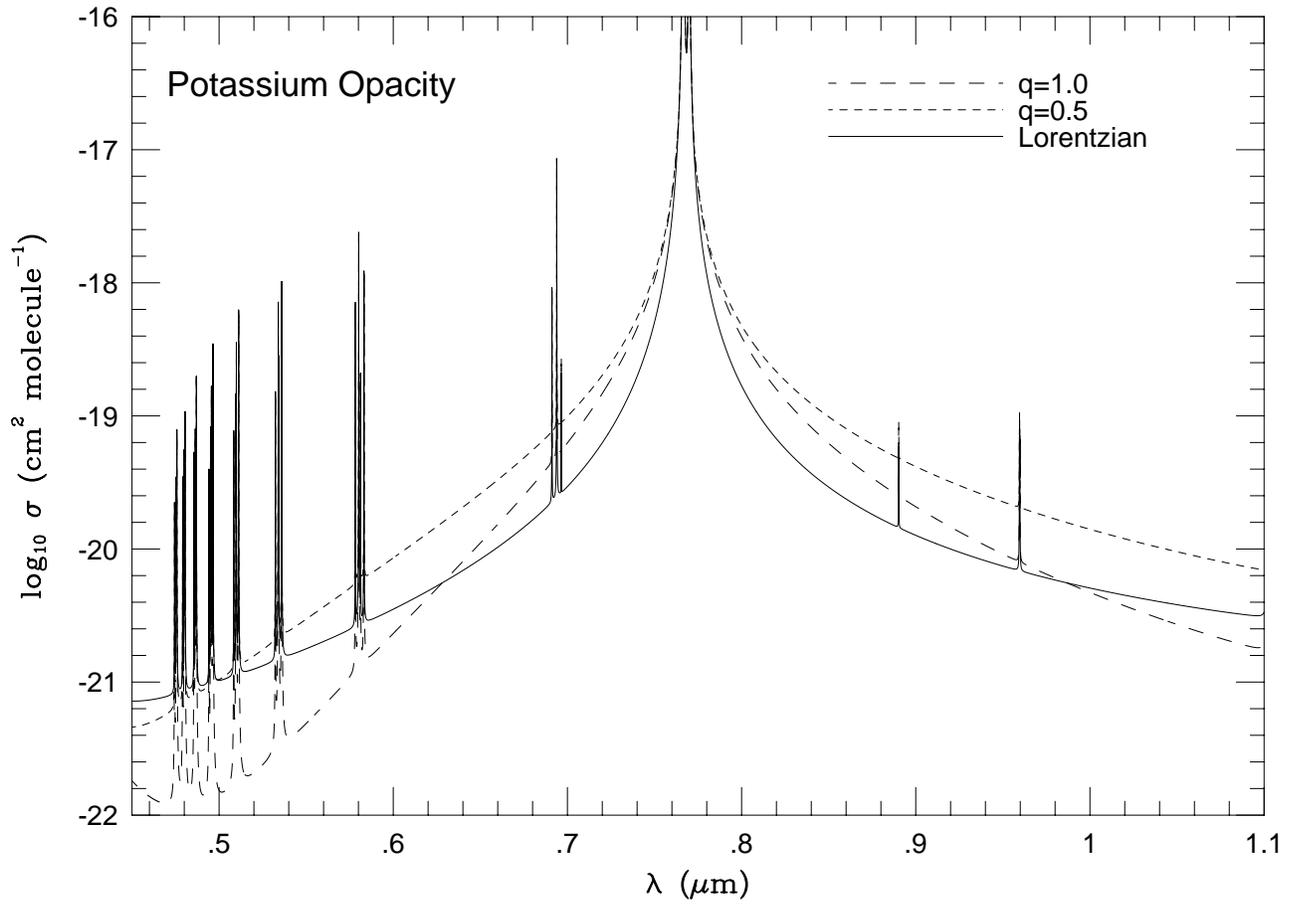}
\caption{
The log of the photon--potassium cross sections per particle, in cm$^2$, versus
wavelength ($\lambda$) in \mic. The focus is on the region around the K I resonance doublet at 7700 \AA\
and for specificity a temperature and pressure of 2000 K and 1 bar have been assumed.
The solid line is the Lorentzian theory without corrections in the broad wings at large detunings.
The short-dashed and long-dashed lines are for the corrected theory, with $q$s of 0.5 and 1.0, respectively.
The crucial fact is the useful slope between 0.8 \mic and 1.0 \mic.
\label{fig:5}}
\end{figure}

\begin{figure}
\epsscale{1.05}
\plotone{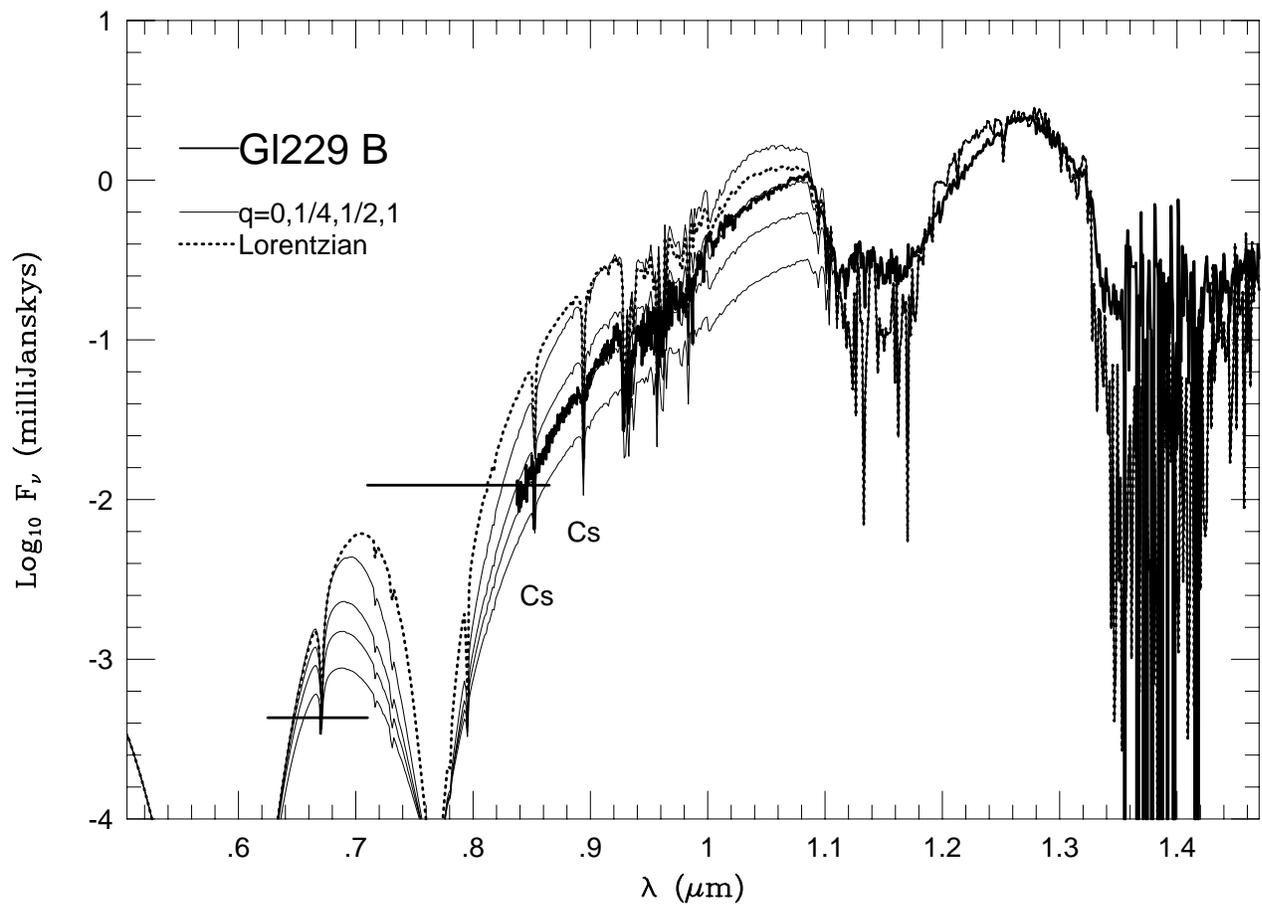}
\caption{
Similar to Figure 1, using the \teff=950 K and $g=10^5$ cm s$^{-2}$ Gliese 229B model,
but for $q$s of 0.0, 0.25, 0.5, and 1.0.  The lower the $q$, the lower the curve.
The dotted line is the theoretical spectrum using the Lorentzian profile.
Alkali metal abundances of $0.3\times$ solar were assumed.  The heavy solid line is the measured spectrum of Gliese 229B
and the solid horizontal lines are the WFPC2 Gliese 229B $R$ and $I$ fluxes.
\label{fig:6}}
\end{figure}

\begin{figure}
\epsscale{1.05}
\plotone{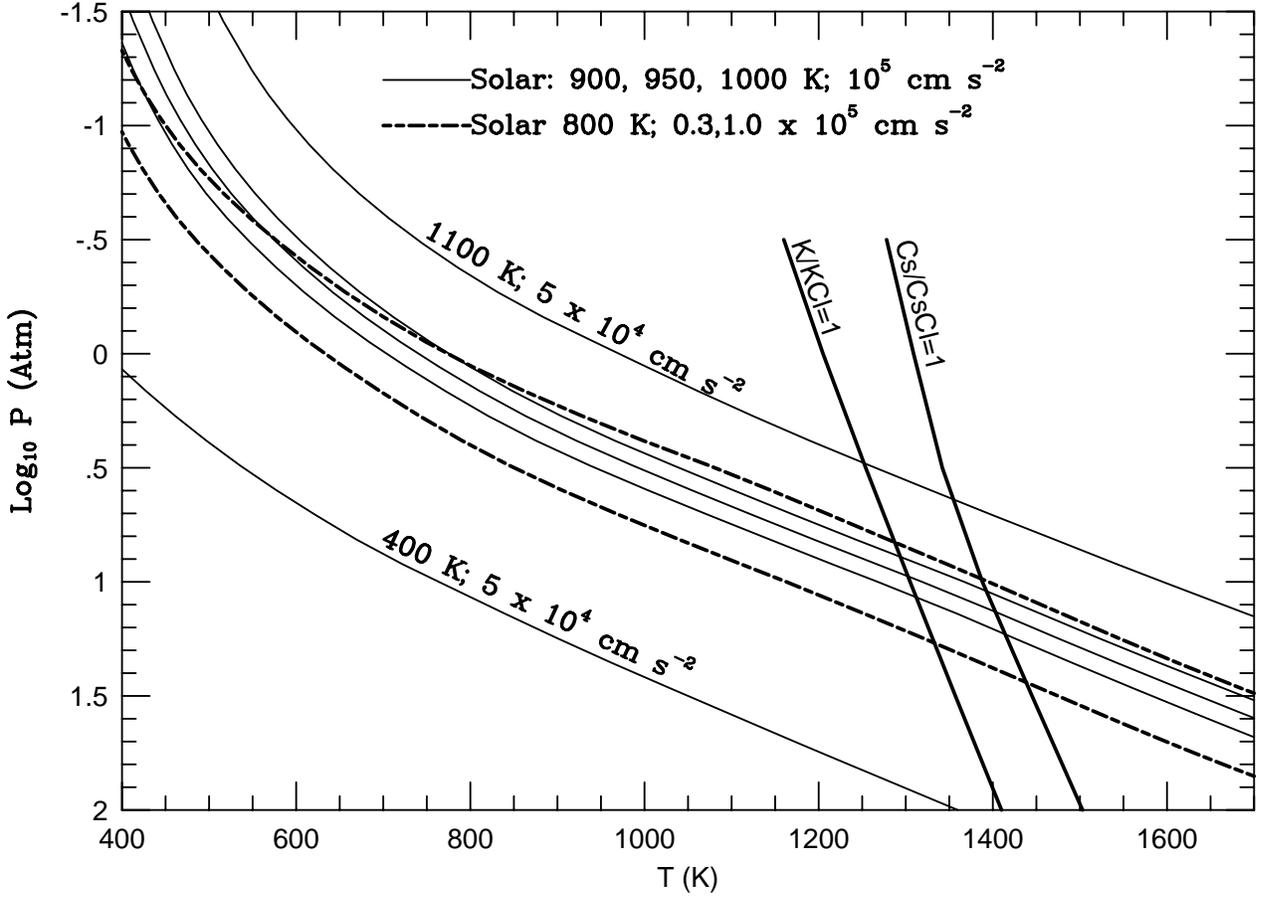}
\caption{
The pressure (in Atm) versus the temperature (in Kelvin) for various
brown dwarf atmospheres, as a function of \teff and gravity and at solar metallicity.  The three central
solid lines are for 900 K, 950 K, and 1000 K models with a gravity of $10^5$ cm s$^{-2}$. The two outliers
are models at temperatures (1100 K and 400 K) and entropies that are too high and too
low, respectively, to fit the Gliese 229 B spectral data.  The long-dash/short-dash curves are models with \teff= 800 K,
the higher one at $3\times10^4$ cm s$^{-2}$ fits, while the lower one at $10^5$ cm s$^{-2}$ does not.
Also included are the lines that demark the true-equilibrium, no-rainout temperature/pressure trajectories for which
atomic K and gaseous KCl and atomic Cs and gaseous CsCl have equal abundances.
While the Cs/CsCl line will be little changed by rainout, the K/KCl line will be shifted to lower temperatures
by perhaps 200--300 K.  These lines serve to indicate the character of the burial of the
elemental alkali metals with decreasing core entropy.  As the core entropy decreases, the pressures
and column depths at the intersection points steadily increase.
\label{fig:7}}
\end{figure}

\begin{figure}
\epsscale{1.05}
\plotone{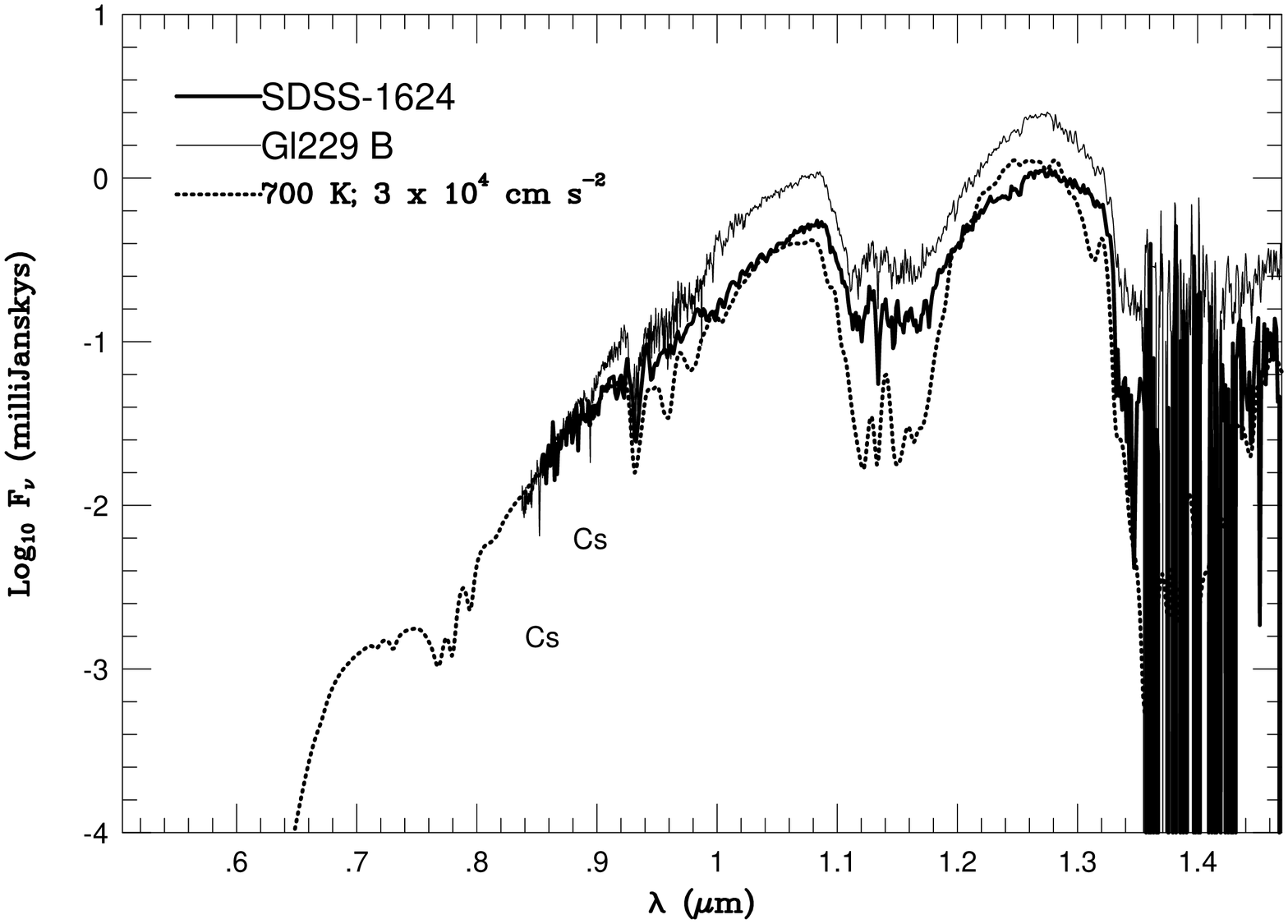}
\caption{
The log of the absolute flux (F$_{\nu}$) in milliJanskys versus wavelength ($\lambda$) in microns
for the first Sloan T dwarf, SDSS 1624+00 (heavy solid line), from Strauss \etal (1999).  The light solid
line is Gliese 229B, while the dotted line is for a model with \teff=700 K and $g=3\times 10^4$ cm s$^{-2}$.
The core entropy of this theoretical model is lower than that for those that fit
Gliese 229B.  Both the model and the Sloan data have a shallower slope than Gliese 229B from 0.85 \mic to 1.0 \mic
and weak cesium lines.
\label{fig:8}}
\end{figure}


\begin{references}

\reference{ } Allard, F., Hauschildt, P.H., Baraffe, I. \& Chabrier, G. 1996,
\apj, 465, L123
\reference{anders89} Anders, E. \& Grevesse, N. 1989, Geochim. Cosmochim. Acta, 53, 197
\reference{ } Anderson, P.W. 1950, Phys. Rev., 79, 132
\reference{} Breene, R.G. Jr. 1957, Rev. Mod. Phys., 29, 94
\reference{} Breene, R.G., Jr. 1981, ``Theories of Spectral Line Shape," (New York: John Wiley), p. 344.
\reference{ } Burgasser, A.J. \etal 1999, accepted to \apj, astro-ph/9907019
\reference{b+95} Burrows, A., Saumon, D., Guillot, T., Hubbard, W.B., \& Lunine, J.I. 1995,
                Nature, 375, 299
\reference{ } Burrows, A., Marley M., Hubbard, W.B. Lunine, J.I., Guillot, T., Saumon, D.
Freedman, R., Sudarsky, D. \& Sharp, C.M. 1997, \apj, 491, 856
\reference{ } Burrows, A. and Sharp, C.M. 1999, \apj, 512, 843
\reference{ } Charbonneau, D. \etal 1999, \apj, 522, L145
\reference{} Ch'en, S. and Takeo, M. 1957, Rev. Mod. Phys., 29, 20
\reference{delf97} Delfosse, X., Tinney, C.G., Forveille, T., Epchtein, N., Bertin, E., Borsenberger, J.,
        Copet, E., De Batz, B., Fouqu\'{e}, P., Kimeswenger, S., Le Bertre, T., Lacombe, F., Rouan, D., \&
        Tiph\`{e}ne, D. 1997, \aa, 327, L25 
\reference{ } Cuby, J.G., Saracco, P., Moorwood, A.F.M., D'Odorico, S., Lidman, C., Comeron, F.,
and Spyromilio, J. 1999, astro-ph/9907028
\reference{ } Dimitrijevi\'{c}, M.S. \& Peach, G. 1990, \aap, 236, 261
\reference{fl96} Fegley, B. \& Lodders, K. 1996, \apj, 472, L37
\reference{ } Geballe, T.R., Kulkarni, S.R., Woodward, C.E., and Sloan, G.C. 1996, \apj, 467, L101
\reference{Go98}Golimowski, D. A., \etal 1998, \aj, 115, 2579
\reference{} Griem, H.R. 1964, ``Plasma Spectroscopy," (New York: McGraw Hill).
\reference{ } Griffith, C.A., Yelle, R.V., and Marley, M.S. 1998, Science, 282, 2063 
\reference{} Holstein, T. 1950, Phys. Rev., 79, 744
\reference{} Holtzmark, J. 1925, Z. f\"ur Physik, 34, 722
\reference{hub77} Hubbard, W. B.\ 1977, Icarus, 30, 305 
\reference{jones97} Jones, H.R.A. \& Tsuji, T. 1997, \apj, 480, L39
\reference{ } Khare, B.N. and Sagan, C. 1984, Icarus, 60, 127
\reference{kirk99} Kirkpatrick, J.D., Reid, I.N., Liebert, J., Cutri, R.M., Nelson, B.,
                   Beichman, C.A., Dahn, C.C., Monet, D.G., Gizis, J., and Skrutskie, M.F. 1999,
                   \apj, 519, 802
\reference{ } Leggett, S., Toomey, D.W., Geballe, T., and Brown, R.H. 1999, \apj, in press
\reference{ } Lodders, K. 1999, \apj, 519, 793
\reference{ } Marley, M.S., Saumon, D., Guillot, T., Freedman, R.S., Hubbard, W.B.,
Burrows, A. \& Lunine, J.I. 1996, Science, 272, 1919
\reference{marley97} Marley, M.S. 1998, in the proceedings of the Tenerife Workshop 
on Extrasolar Planets and Brown Dwarfs, ed. R. Rebolo,
E.L. Martin, \& M.R. Zapatero--Osorio (ASP Conf. Series V. 134), p. 383.  
\reference{ } Marley, M.S., Gelino, C., Stephens, D., Lunine, J.I., \& Freedman, R. 1998, \apj, 513, 879
\reference{} Matthews, K., Nakajima, T., Kulkarni, S.R., \& Oppenheimer, B.R. 1996, \aj, 112, 1678
\reference{ } Nakajima, T., Oppenheimer, B.R., Kulkarni, S.R., Golimowski, D.A.,
Matthews, K. \& Durrance, S.T. 1995, Nature, 378, 463
\reference{ } Nefedov, A.P., Sinel'shchikov, V.A., and Usachev, A.D. Physica Scripta, 59, 432
\reference{ } Noll, K., Geballe, T.R., \& Marley, M.S. 1997, \apj, 489, 87
\reference{opp95} Oppenheimer, B.R., Kulkarni, S.R., Matthews, K., \& Nakajima, T. 1995,
                  Science, 270, 1478
\reference{opp98} Oppenheimer, B.R., Kulkarni, S.R., Matthews, K., \& van Kerkwijk, M.H. 1998,
                  \apj, 502, 932
\reference{ } Piskunov, N.E., Kupka, F., Ryabchikova, T.A., Weiss, W.W.,
\& Jeffery, C.S. 1995, \aap Suppl., 112, 525
\reference{saumon96} Saumon, D., Hubbard, W.B., Burrows,
A., Guillot, T., Lunine, J.I., \& Chabrier, G. 1996, \apj, 460, 993
\reference{ } Schultz, A.B. \etal 1998, \apj, 492, L181
\reference{ } Seager, S. and Sasselov, D.D. 1998, \apj, 502, L157
\reference{ } Spitzer, L., Jr. 1940, Phys. Rev., 58, 348
\reference{ } Strauss, M.A. \etal 1999, submitted to \apj, astro-ph/9905391
\reference{ } Sudarsky, D., Burrows, A., and Pinto, P. 1999, in preparation
\reference{ } Tinney, C.G., Delfosse, X., Forveille, T. and Allard, F. 1998, \aap, 338, 1066
\reference{tsuji96} Tsuji, T., Ohnaka, K., Aoki, W., \& Nakajima, T. 1996, \aa, 308, L29
\reference{tsuji99} Tsuji, T., Ohnaka, K., and Aoki, W. 1999, \apj, 520, L119
\reference{ } Tsvetanov, Z.I. \etal 1999, submitted to \apj
\reference{ } Uns\"old, A. 1955, ``Physik der Sternatmosph\"aren," 2'nd edition (Berlin: Springer-Verlag), p. 305
\reference{ } Weisskopf, V. 1933, Z. f\"ur Physik, 43, 1

\end{references}
\end{document}